\documentclass[twocolumn,pre]{revtex4}
\usepackage{color}
\usepackage{morefloats}
\usepackage{amssymb}
\usepackage{amsmath}
\usepackage{float}

\usepackage{graphicx} 

\begin{document}

\newcommand{\ie}{i.~e.~}

\title{Investigation of tracer diffusion in crowded cylindrical channel}
\author{Rajarshi Chakrabarti, Stefan Kesselheim, Peter Ko\v{s}ovan and Christian Holm}

\affiliation{Institut f\"ur Computer Physik, Universit\"at Stuttgart, Pfaffenwaldring
27, 70569 Stuttgart, Germany. E-mail: holm@icp.uni-stuttgart.de}

\begin{abstract}
  Based on a coarse-grained model, we carry out molecular dynamics
  simulations to analyze the diffusion of a small tracer particle
  inside a cylindrical channel whose inner wall is covered with
  randomly grafted short polymeric chains. We observe an interesting
  transient subdiffusive behavior along the cylindrical axis at high
  attraction between the tracer and the chains, however, the long time
  diffusion is always normal. This process is found to be enhanced for
  the case that we immobilize the grafted chains, i.e. the
  sub-diffusive behavior sets in at an earlier time and spans over a
  longer time period before becoming diffusive. Even if the grafted
  chains are replaced with a frozen sea of repulsive, non-connected
  particles in the background, the transient subdiffusion is
  observed. The intermediate subdiffusive behavior only disappears
  when the grafted chains are replaced with a mobile background sea of
  mutually repulsive particles. Overall, the long time diffusion
  coefficient of the tracer along the cylinder axis decreases with the
  increase in system volume fraction, strength of attraction between
  the tracer and the background and also on freezing the background.
  We believe that the simple model presented here could be useful for
  a qualitative understanding of the process of macromolecular
  diffusion inside the nuclear pore complex.
\end{abstract}

\maketitle
\section{Introduction}

Diffusion in crowded environment has been an active area of
experimental and theoretical research \cite{sokolovsoftmatter, pasquali, fytas,
fytaslangmuir, hofling2011, weitz, metzlerprl2011, frey, greene, metzlerpnas,
schreiber}. Especially in biology it is not uncommon to find aqueous
crowded environments with agents that are important for biological functions.
Many examples where crowding and/or sticky interaction between the diffusing 
species and the surroundings changes the rate and even the nature of the
diffusion process have been found in biological or synthetic contexts:

DNA-binding proteins search
for specific binding sites on a DNA molecule by combining one dimensional
diffusion along the DNA strand and three dimensional diffusion in the bulk
\cite{greene, metzlerpnas}. 
The self diffusivity of proteins in 
crowded solutions has been shown to be slowed down to one fifth of the dilute
limiting value \cite{schreiber}. 
Also diffusion of proteins across the nuclear pore complex (NPC)
\cite{bayliss, alberts, raminprl, schulten2, bird, Zilman2007, dekker}
is an example of diffusion in a crowded environment where the crowding
is due to the presence of proteins called nucleoporins that are rich
in hydrophobic amino acids and form a brush \cite{ Frey2007, elbaum}
or a reversible hydrogel \cite{Lim, elbaum}. Proteins diffusing
through NPC bind to these nucleoporins and experience a slowdown.  A
thorough explanation of this effect would lead to a better
understanding of the biological functioning of the NPC \cite{frey,
  Lim}.

Another issue is the emergence of non-Fickian, anomalous diffusion in presence
of crowding \cite{sokolovsoftmatter}: Evidence of anomalous diffusion was found
for  colloidal tracer particles in entangled actin filament networks
\cite{weitz} and for soluble proteins in highly viscoelastic cytoplasm and
nucleoplasm \cite{weissfebslett}.  Very recently it has been shown that
diffusion inside the NPC could also be anomalous \cite{liphardt}.  On the other
hand sticky interaction can also lead to subdiffusive behavior even in absence
of crowding. It has been shown experimentally  that functionalized colloidal
particles with DNA sticky ends diffuse anomalously on a complementarily coated
surface \cite{chaikin}.  Diffusion in hydrogels and gel-like media
\cite{holmtracer, amsden, fytasjpcb, fytas, schweizerjpcb2008} are also
examples of diffusion in crowded environment and subdiffusive behavior in an
intermediate time scale has been observed in computer simulation
\cite{holmtracer} and signatures of subdiffusive behavior has been observed
experimentally \cite{fytas}.

These observations suggest that similar physical principles govern the diffusion
processes in under very different conditions. We investigate these principles 
in a simple and generic model and shed light
the role of crowding and sticky interaction with means of molecular dynamics 
simulations. 


We study the diffusion of
tracer particles inside a cylindrical channel grafted with polymeric chains
from the inside. The polymeric chains consists of mutually repulsive monomers connected with 
springs. In the field of polymer science, this is a typical model for a polymer in a good solvent \cite{grest86a}.
Tracers are added to the system and the impact of their interaction with
the polymer chain on their diffusional property is studied. The thermal effects of the surrounding water is replaced by 
a standard LAngevin thermostat. In this
sense it is a solvent--implicit model where hydrodynamic effects are neglected.

From a  biological perspective this model, although being very simple, is 
inspired by the brush model of the
central plug of NPC \cite{Patel2007}. Nucleoporins rich in hydrophobic
units form a brush--like structure
\cite{Lim} and it is believed that proteins being  transported through the
central plug bind to these nucleoporins \cite{Frey2007, herrmann}.  These
bindings are believed to be hydrophobic in nature, and  each binding is in the range of
$1-2$ $k_B T$ \cite{mofrad, mofradplos}.  


To mimic the aspect of crowding we chose a repulsive interaction
between all chain monomers
and the tracer particles. To investigate the role of interaction we incorporate 
a Lennard--Jones potential between the tracer and the monomers of which we 
vary the interaction strength in a range from $1-2.5$ $k_B T$ corresponding
to the interaction strength of hydrophobic contacts between proteins and
nucleoporins in the NPC. The resulting diffusive behaviour is characterized
by investigating the mean--square--displacement for the tracer on different timescales.



The paper is structured as follows. In section II, the model, the
simulation methods and the classification of observed diffusion
processes are discussed. In Section IV the results and interpretations
for the different models are given. The paper ends with the
conclusions in section IV.

\section{Model and Method}


The model described in the following section is evaluated using molecular
dynamics. We use a generic unit system in which the
Lennard--Jones parameter $\sigma$ is chosen as the unit of length, the
temperature multiplied with Boltzmann's constant as the unit of energy and mass
is measured in units of the mass of the monomers in the chain molecules. All quantities in
the following are expressed in this unit system.

A cylinder with height $24\sigma$ and radius $9\sigma$ is created from
particles with a diameter $\sigma$ and a mutual distance of 1 $\sigma$ so
that a closed cylindrical surface is formed on which each particle has
four next neighbors. These particles, we call them wall particles, are fixed in
space, \ie their equations of motions are not integrated, independent of the
force acting on them.  The cylinder axis is chosen to be the $z$ axis.  
 Each of the polymeric chains is made of $N=13$  monomers, connected through a finite
extensible nonlinear elastic (FENE) potential:
\begin{equation}
U_{fene}(r)=-\frac{K_f r_{max}^2}{2}log\left[1-(r/r_{max})^2\right]\label{fene}.
\end{equation}
Here $K_f$ is the force constant and $r_{max}$ is the maximum displacement of the bond. 
In the simulation the parameters are chosen to be $K_f=7$, $r_{max}=2$, $N=13$.
The cylindrical surface is grafted randomly with
polymeric chains by fixing the end monomer of each of the chains to
one of the wall particles.

The repulsive interaction between the monomers themselves and with the
wall is  modeled via the purely
repulsive variant of the Lennard-Jones potential also known as the Weeks-Chandler-Andersen
(WCA) potential \cite{weeks71a}: 
\begin{equation}
V_{WCA}(r) =
\begin{cases}
4\epsilon \left[(\frac{\sigma}{r})^{12}-(\frac{\sigma}{r})^{6}\right]+\epsilon & ,\text{if }r<(2)^{1/6}\sigma \\
0 & ,\text{otherwise},
\end{cases}
\label{WCA}
\end{equation}
with $\epsilon=1$ and $\sigma=1$.

The tracer particles are of the 
same size $\sigma$ and of the same mass as the monomers. They interact
with the wall particles by means of the WCA potential, and with the
grafted polymers either via the WCA potential (for simple crowding), or by an
attractive (=``usual'') Lennard-Jones potential:
\begin{equation}
V_{LJ}(r) =
\begin{cases}
4\epsilon \left[(\frac{\sigma}{r})^{12}-(\frac{\sigma}{r})^{6}\right] & ,\text{if }r<r_\text{cut} \\
0 & ,\text{otherwise},
\end{cases}
\label{lj}
\end{equation}
where the cut off radius is fixed at $r_\text{cut}=2.5$ and the attraction
strength $\epsilon$ is varied.  

The degree of crowding is characterized by the grafting density $\gamma$, \ie
the number of chains per unit are of the cylinder.
A typical value of $n=75$ chains
corresponds to $\gamma = 0.055$. 

\begin{figure}
\centering
 \includegraphics[width=0.4\textwidth]{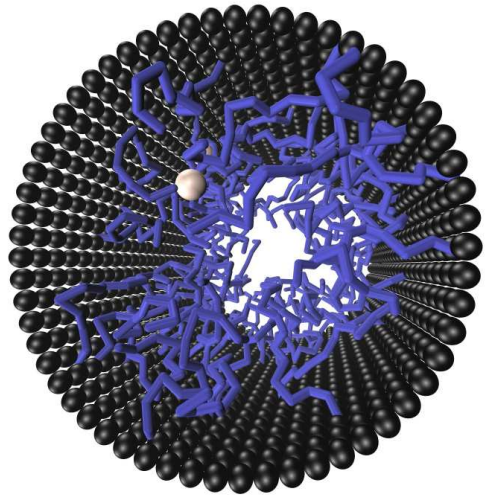}
 \caption{Illustration of the simulation set up. Tracer particles (white
 spheres) inside a cylindrical pore with rigid walls (black) grafted from
 inside with polymeric chains (blue). For graphical clarity the polymer chains
 are represented as (thinner) continuous curves (VMDs \emph{licorice} drawing method),
 and the size of the tracer particles is exaggerated. 
}
\label{fig:snapshot}
\end{figure}

A typical representation of the investigated system is shown in Fig.
\ref{fig:snapshot} by using  VMD \cite{vmd} . 

We performed five independent simulations for each data point with 2
million timesteps each in order to allow to estimate the statistical
errors indicated in all figures.  All simulations are carried out with
the Langevin thermostat, thus solving the full phase space stochastic
equation of motion known as the Langevin equation of motion. Periodic
boundary conditions along the cylindrical axis are used throughout the
simulation and the velocity Verlet algorithm is used with a time step
of $\tau=0.005$.

For a particle with mass $m$ and the position co-ordinate $r_i$ (where
$i$ stands for the $i$the particle), interacting with all other
particles ($i\neq j$) through a general potential $U(r_{ij})$ the full
phase space (underdamped) Langevin equation of motion reads as
\begin{equation}
  m\frac{d^{2}\mathbf{r}_{i}(t)}{dt^{2}}=-\nabla_{r_i}\sum_{j\neq i}U(\mathbf{r}_{ij})+\mathbf{F}_{i}^{D}+\mathbf{F}_{i}^{R}\label{langevin}
\end{equation}
where 
\begin{equation}
\mathbf{F}_{i}^{D}=-m\Gamma\frac{d\mathbf{r}_{i}(t)}{dt}
\end{equation}
is the frictional drag acting on the particle with friction coefficient $\Gamma$,  $F_{i}^{R}$ is the random force acting on the particle.
The first moment of the random force is chosen to be zero:
\begin{equation}
\langle\mathbf{F}_{i}^{R}(t)\rangle =0.
\end{equation}
The magnitude of the random forces obeys
\begin{equation}
\langle\mathbf{F}_{i}^{R}(t_1)\mathbf{F}_{j}^{R}(t_2)\rangle =6\Gamma k_B T m \delta_{ij}\delta(t_1-t_2)
\end{equation}
where $k_B$ is the Boltzmann constant and $T$ is the temperature of the Langevin thermostat.

The delta function in time between the random forces in the above
equation ensures the spectrum of the random force to be white, corresponding to a Markovian process in phase space with no memory. 
\emph{What is that: This also means dissipation is Ohmic}. Throughout the simulation hydrodynamic 
interactions are neglected. 
With the above choice of random forces a fluctuation--dissipation theorems, that connects the diffusion coefficient $D$ and
the friction coefficient $\Gamma$ holds: The diffusion coefficient of
a single particle, up to a factor $k_B T/m$ is the inverse of
the friction coefficient:
$D_\text{free}=\frac{k_B T}{m \Gamma}=1$. 
In all simulations we apply a friction coefficient of $1$ and the diffusion
coefficients reported in the paper are given in units of $D_\text{free}$, 
being also 1 in our unit system.

The molecular dynamics simulation have been carried out with the help of the
simulation package ESPResSo \cite{holmespresso} in version 3.1 \cite{arnold12a}. 

To describe the diffusion of tracer particles it is a common practice to use
the mean square displacement (MSD), $\langle \Delta \vec r_{t_0}\left(t\right) ^2\rangle$ as a
measure, where $\Delta \vec r(t)=\left(\vec r(t+t_{0})-\vec r(t_0)\right)$ is the
displacement of the particle at time $t+t_0$ from its initial position at $t_0$.
The angular bracket denotes an average over all possible realizations of the
process. Under the time translational
invariance this quantity is independent of $t_0$ and this index can be dropped,
thus $t$ can also be interpreted as the lag time. For ergodic systems,  
the average over realizations can be replaced by an ensemble average. In case of normal diffusion,
MSD increases linearly with time and in $d$ dimension reads as $\left< \Delta
\vec r(t)^2\right>=2 d D t$, where $D$ is the diffusion coefficient of the
tracer particle. 
On the other hand, a process
where the MSD is not linear in time but proportional to $t^{\alpha}$ with $\alpha
\neq 1$ is referred to as anomalous or non-Fickian
diffusion \cite{sokolovsoftmatter}.  Here $\alpha$ is the diffusion exponent
and if $\alpha<1$, the process is called subdiffusive and if $\alpha>1$, the
process is called superdiffusive. For anomalous diffusion there is no well
defined diffusion coefficient. While the definition given above is very helpful for
isotropic environments, in anisotropic environments the diffusion constant
needs to be replaced by a diffusion tensor. As we are only interested in the
diffusive behaviour along the cylinder axis (=$z$--axis), we measure
the MSD in $z$ direction:
\begin{equation}
  \text{MSD} = \langle \left[r_z\left(t + t_{0}\right)-r_z\left(t_{0}\right)\right]^2 \rangle,
  \label{MSD}
\end{equation}
corresponding to the $zz$ component of the diffusion tensor. Throughout the paper $D$ always
represents the tracer diffusion coefficient along the cylinder axis and
MSD is the mean square displacement of the tracer along the cylinder axis.

  








\section{Results}

In this section, we present the results of the MD simulations.  We
compute the MSD of the tracer and analyze the resulting transient
diffusive behaviour with respect to the question if anomalous
diffusion appears and try to elucidate the reasons for this by
comparison to reference systems.

To explore the role of chain connectivity we compare all our results to a system
when the chain monomers are not connected via the FENE potential, but free to move thus forming a gas or a ''sea`` of
independent particles. The other reference is a system in which the grafted
chains are moving much slower than the tracer particle, we call this a frozen
background. For brevity we characterize the four possible combinations  by the keywords pairs
\emph{(chains/particles)} and \emph{(mobile/frozen)}.
We will show that both properties, the chain connectivity and the
background motion, have a distinct impact on the diffusion properties of the
tracer particles.

\begin{figure}[h]
\flushleft
\hspace{-0.9cm}
 \includegraphics[width=\columnwidth]{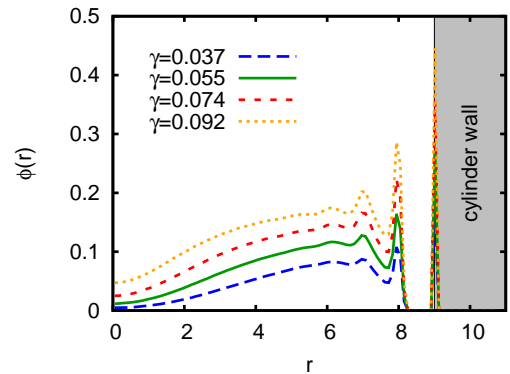}
 \caption{The local monomer volume fraction, $\phi(r)$ at different chain grafting densities ($\gamma$) of the mobile grafted chains.
 Increasing the number of chains only affects the density but not the structure of the brush.}
\label{fig:rhomandgamma}
\end{figure}

First we discuss the structure of the polymer coating: The radius of gyration
of the chains is $\simeq 6$ under all reported conditions and the
mutual distance between grafting points
is smaller also for the smallest grafting densities. Thus we are in the regime
where the grafted chains overlap considerably. In Fig.~\ref{fig:rhomandgamma} 
we report the local volume fractions of chain monomers, i.e. the fraction
of the volume locally occupied by monomers, assuming they are spheres of diamter $\sigma$.
The local volume fractions of up to 10\% and more also indicate that locally
the crowding is significant: Under these conditions mean free paths $~\sim \sigma$
are to be expected.

\subsection{Repulsive Tracer--Chain Interaction}

\begin{figure}[h]
\begin{center}
    \includegraphics[width=0.9\columnwidth]{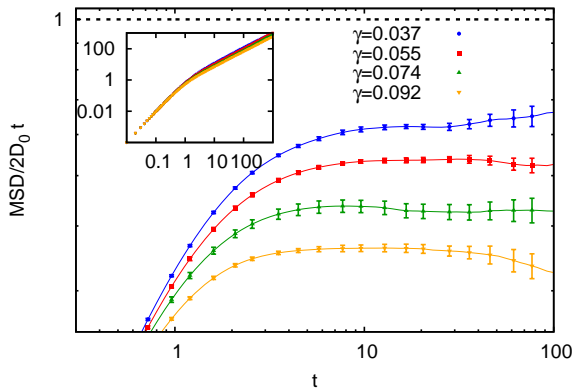}
 \end{center}
 \caption{Comparison of $\mathrm{MSD}/2D_{0}t$  of the purely repulsive tracer along the cylinder axis of the cylinder with mobile grafted chains in the background at different chain grafting densities ($\gamma$). The diffusion constant (seen as the $t\rightarrow\infty$ limit of the curves) depends strongly on the density of grafted chains.
 }
  \label{fig:MSDby2dt_repulsion_poly_mobile}
\end{figure}

We now discuss the situation where the interaction between tracer
particles and grafted chains is purely repulsive.  We perform simulations with
different numbers of grafted chains ($n$) hence different grafting densities
($\gamma$) and calculate the mean square displacement (MSD). We investigate four
grafting densities seen in Fig.~\ref{fig:MSDby2dt_repulsion_poly_mobile}. In
the double-logarithmic plot of the MSD, all lines nearly collapse (seen in the
inset of Fig.~\ref{fig:MSDby2dt_repulsion_poly_mobile}). We thus plot the MSD
of the tracer along the cylinder axis divided by $2D_0t$, where $D_0(=
D_\text{free})$ is the free diffusion coefficient along the cylinder axis. 
Plotting this \emph{reduced MSD} allows to see the
different characteristics of diffusion.  On the timescale $\frac{1}{\Gamma}$ a
ballistic regime where the MSD is proportional to the square of $t$ is
observed, corresponding to a linear increase of the reduced MSD, and on longer
timescales the diffusion becomes normal.  As depicted in
Fig.~\ref{fig:MSDby2dt_repulsion_poly_mobile} the mean square displacement of
the tracer particles shows no sign of anomalous diffusion for all investigated
grafting densities.

\begin{figure}
\centering
 \includegraphics[width=0.4\textwidth]{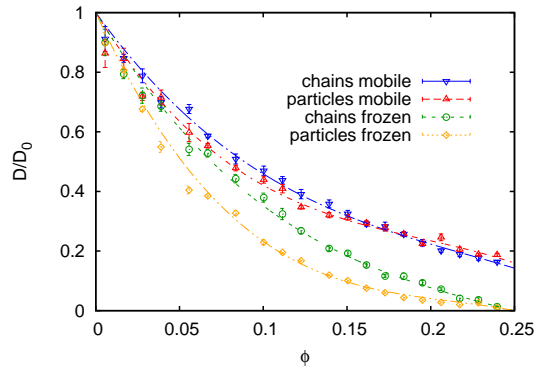}
 \caption{The relative diffusion coefficient $D/D_{0}$ for the purely repulsive tracers against the volume fraction of the system. The lines are guides to eyes. We compare the case where the monomers are connected by springs to form a \emph{chain} to the case where they are
 isolated \emph{particles} and the case where they are \emph{mobile} vs.~the case where they are \emph{frozen}.}
\label{fig:diffusion repulsion}
\end{figure}

The long time diffusion coefficient is reduced compared to a freely diffusing
particle by collisions with the polymer chains. For low volume fractions of
polymer chains we thus expect the diffusion coefficient to be reduced by a
factor proportional to the (average) volume fraction, thus the probability to collide
with a monomer per unit time.  Neither the chain connectivity nor the question
if the background moves on the same velocity scale as the tracer particles
significantly influences this effect.  In Fig.~\ref{fig:diffusion repulsion} we
report the dependence of the diffusion coefficient on the monomer volume
fraction, $\phi=\frac{\frac{4}{3} \pi (\sigma/2)^3 N n}{\pi {r_c}^2 L_c}$. In
all cases the tracer diffusion coefficient decreases similarly with increasing
monomer volume fraction. If the monomers are connected and form chains, the
impact on the diffusion coefficient is lower, as the volume from which the
tracer is excluded is smaller. Also notice that the diffusion becomes
faster with a mobile
background than with a frozen background. This is because in the frozen
background the tracer practically collides with an infinite mass and
is more
likely to get bounced back to its initial position. At higher volume fractions
the topology of the background becomes less important. As can be seen from the
Fig.~\ref{fig:diffusion repulsion}, the two data sets corresponding to frozen
particles and chains coincide at higher volume fractions  as do the data
sets for mobile particles and chains.

\subsection{Attractive Tracer-Chain Interaction}
The dynamics of a tracer particle which interacts with the chains through an
attractive Lennard-Jones potential as defined in (Eq.(\ref{lj})) is more
complex and shows interesting features that are missing in the case of a purely
repulsive interaction.  Very different modes of motion are important in case of
attractive polymer--tracer interaction: An attractive tracer can get attached
to one of the chains, follow the chains movement, move along the chain, hop
from one chain to the other or hop back to the same chain. It can also move back
to the bulk where it executes normal diffusion. The emerging process is a
combination of all these modes of transport and should strongly depend on the
strength of attraction between the tracer and the chain and also on the density
of the grafted chains. We do not expect anomalous diffusion on
long time scales, but see this rather as a transient phenomenon. This
is what one would
expect as long as the noise is white and the background is translationally invariant.
In the simulations the strength of attraction between the tracer and the
polymer ($\epsilon$) is varied from $1$ to $2.5$ .  The
choice of this range of attraction is inspired by the fact that during the
transport through NPC proteins bind to the nucleoporins through hydrophobic
contacts which are in the range of $1-2$ $k_B T$. \cite{mofradplos}. 



\begin{figure}[h]
 \flushleft 
 \includegraphics[width=\columnwidth]{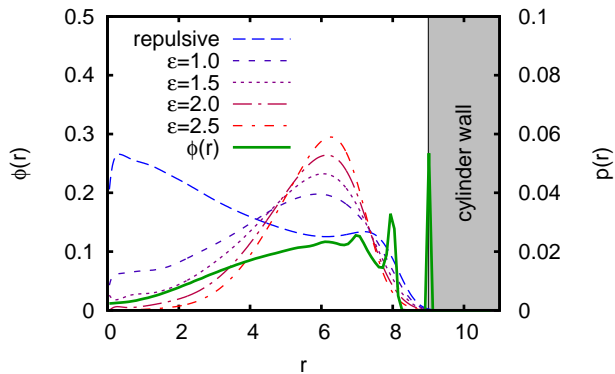}
 \caption{The probability of finding the tracer along the radial direction ($p(r)$) for different attraction strengths $\epsilon$ at 
 fixed grafting density  $\gamma=0.055$. For comparison we repeat the
 density of monomers $\phi(r)$ from Fig.~\ref{fig:rhomandgamma} (solid line).
 For repulsive chain--tracer interactions the tracer is pushed towards the polymer--free center of the cylinder, while the
 attractive interaction pulls it into the polymer brush.
 }
\label{fig:rhomandrhot}
\end{figure}

First we investigate the influence of the attractive interaction on the position where the tracer can be found. In Fig.~\ref{fig:rhomandrhot} the probability density
to find the tracer at a given distance $r$ from the cylinder axis for different attraction
strength at a fixed grafting density 0.055 is reported.
Notice that the probability of finding the tracer along the radial direction
($p(r)$) changes profoundly on making the tracer--chain interaction attractive.  In case of purely repulsive tracers, $p(r)$ is high at the center
of the cylinder ($r=0$) and gradually decreases towards the chains and vanishes
at the cylinder wall ($r=9$). On the other hand in case of attractive tracers
the density profile is qualitatively different. The tracer has a low
density around the center of the cylinder ($r=0$) and has a high density
close to the chains. As expected the attractive interaction significantly 
pushes the tracers from the center of the cylinder towards the area of
the grafted chains. For comparison we also insert the density profile
$\phi(r)$ of chain monomers.

\begin{figure}[h]
  \begin{center}
    \includegraphics[width=0.9\columnwidth]{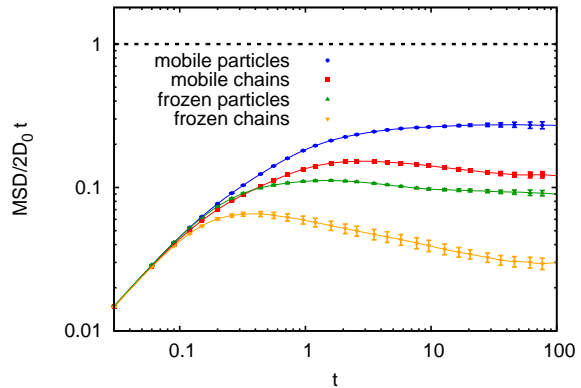}
  \end{center}
  \caption{Comparison of $\mathrm{MSD}/2D_{0}t$  along the cylinder axis for different cases for $\epsilon=2$ at the grafting density $\gamma=0.055$. Subdiffusive transient behaviour indicated by a negative slope is seen in all cases, except for the case where the monomers are not connected and allowed to move. Chain structure and freezing the background facilitates subdiffusion.}
  \label{fig:msdby2dt_e2_all_n75}
\end{figure}

Increasing the attraction strenght between the tracer particle and the chains
in general slows down the diffusion. In table \ref{tbl:example} we report the
long time diffusion coefficients obtained for different attraction strengths
($\epsilon$) at the grafting density $\gamma=0.055$. When the attraction strength
is increased, a subdiffusive regime for an intermediate time period is observed
in the plot of the reduced MSD against $t$ (Fig.~\ref{fig:msdby2dt_e2_all_n75})
for $\epsilon\geq2 k_B T$.
\begin{table}[h]
\small
  \caption{\ $\frac{D}{D_{0}}$ of the tracer with grafted mobile chains in the background at the grafting density $\gamma=0.055$ for different attraction strengths ($\epsilon$). Increased attraction strength slows down diffusion.}
  \label{tbl:example}
  \begin{tabular*}{0.2\textwidth}{@{\extracolsep{\fill}}ll}
    \hline
   $\frac{D}{D_0}$  & $\epsilon$  \\
    \hline
  1.0  & 0.38 $\pm$ 0.02     \\
  1.5  & 0.23 $\pm$ 0.01     \\
  2.0  & 0.125 $\pm$ 0.005   \\
  2.5  & 0.074 $\pm$ 0.003  \\
    \hline
  \end{tabular*}
\end{table}

We give the following interpretation: The energy landscape seen by a
tracer particle for a fixed configuration of the polymer chain has
minima in which the particle gets trapped on these intermediate
timescales. These minima are especially deep in vicinity of kinks in
the polymer chain, where the attractive regions of several monomers
overlap. This also occurs in vicinity of points, where two chains
closely pass by each other. When a particle gets trapped, it moves
together with the monomers responsible for the trapping. Since the
chains are connected the corresponding displacement will only be
small.  The structure of the polymer chains itself is subject to
Brownian motion, and this motion eventually helps the tracer to escape
the traps.

To justify this interpretation we perform independent control experiments with
a system where the chain monomers are not connected (a sea of \emph{particles})
and investigate the impact of strongly slowing down the motion of the
background, hence look the motion of the tracer particle on a \emph{frozen} background. We
report here only the results for attraction strength $\epsilon=2$, but qualitatively
the results that were obtained for the other investigated attraction strengths of 1, 1.5 and 2.5 $k_B
T$ are similar.

\begin{figure}[h]
  \begin{center}
    \includegraphics[width=0.9\columnwidth]{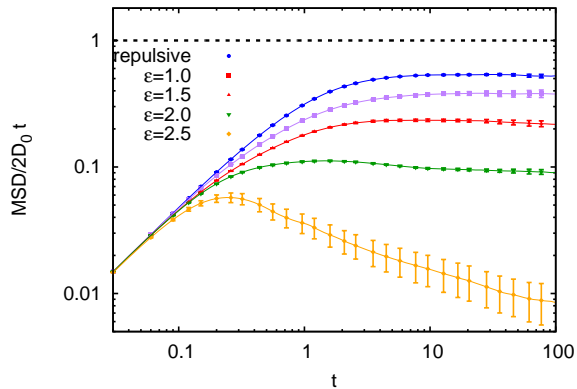}
  \end{center}
  \caption{Comparison of $MSD/2D_{0}t$  along the cylinder axis for the purely repulsive case and different attraction strengths ($\epsilon$) with mobile grafted chains in the background.}
  \label{fig:MSDby2dt_eall_poly_mobile_n75}
\end{figure}

The observed reduced MSDs in all four cases are shown in
Fig.~\ref{fig:MSDby2dt_eall_poly_mobile_n75}.  By giving up the chain structure
of the environment we expect to facilitate diffusion as the deep traps where
the attractive regions of several monomers overlap are less likely as entropy
drives them away from each other. The long time diffusion coefficient is
increased by a factor of $2.2$ and the reduced MSD exhibits no subdiffusive
behaviour any more.  Fixing the position of the disconnected monomers slows
down diffusion again, as this suppresses the joint diffusion of complexes of
tracers and monomers, but merely stops the tracers until they can escape from
their traps. For $\epsilon = 2$ this stopping effect is so pronounced that a
transient subdiffusive behaviour occurs (corresponding to the negative slope of
the green curve in Fig.~\ref{fig:MSDby2dt_eall_poly_mobile_n75}), but for
weaker attraction it disappears.

Freezing the structure of the chains does not change the distribution of traps,
but their lifetimes.  Trapped particles have to escape by their own Brownian
motion and the random motion of the chains does no more facilitate their escape
due to random displacements of the traps.  It is seen that a substantially more
pronounced subdiffusive regime occcurs. The long time diffusion coefficient is
decreased by $70$ $\%$ by freezing the background. This observation
qualitatively supports the proposition of Bickel and Bruinsma \cite{bickel}.
They showed with a very simple model for the transport across NPC that the
fluctuating network of chains in the background acts as an extra noise and
actually enhances the diffusion of a protein immersed in it as compared to in a
frozen network.

\section{Conclusions}

In this work, we have investigated the process of diffusion of tracer
particles inside a crowded cylindrical channel. Our coarse grained
simulations show that if the interaction between the environment and
the tracer particle is purely repulsive, the diffusion is slowed down
significantly: Up to 80\% if the background moves, up to 95\% if the
background is frozen. No subdiffusive behaviour is observed on any
timescale. If the tracer interaction, however, is only weakly
attractive (1-2.5 $k_B T$) we observe a stronger slowdown than in the
repulsive case, and also a pronounced subdiffusive regime. The
subdiffusive regime is enlarged by freezing the background,
corresponding to a much slower motion of the background
particles. These findings are summarized in table
\ref{tbl:conclusion}.

Our findings suggest that the subdiffusive behaviour observed by Lowe
et.~al.\cite{liphardt}~for the diffusion of tracers in the NPC is not
caused by crowding from a polymer brush alone. Our results clearly
demonstrate that simple attractive interactions between the crowding
polymer-brush and the transported particles are sufficient to observe
transient subdiffusive behaviour. The presence of attractive
interactions were recently revealed in the work of Lim et
~al.\cite{lim07a} who could show that proteins that help transporting
cargo through the NPC, do have an attractive interaction with the
nucleoporins polymers in the NPC.  It, however, remained unanswered if
the nucleoporins form a gel rather than a brush phase and what
consequences this has for the cargo transport through the NPC.  In the
future we want to refine and extend our presented model into this
direction.

\begin{table}[h]
\centering
\small
  \caption{No Subdiffusion is observed in case of purely repulsive tracers. Results for attractive tracers are summarized below.}
  \label{tbl:conclusion}
 \begin{tabular}{lp{2cm}p{2cm} }
    \hline
     $Background$ & $Mobile$ & $Frozen$  \\
    \hline
    Chains & Transient subdiffusion at intermediate attraction & Transient subdiffusion at intermediate attraction   \\
    \hline
   Particles & No subdiffusion & Transient subdiffusion at high attraction  \\
    \hline
\end{tabular}
\end{table}

\section{Acknowledgments}

Illuminating discussion with Dr. Olaf Lenz and Dr. Felix H\"{o}fling
are gratefully acknowledged .  We also thank Volkswagenstiftung and
the Collaborative Research Centres 716 (SFB 716) for providing the
necessary financial funds.

\bibliographystyle{apsrev} 

\end{document}